\documentclass[a4paper,11pt]{article}
\usepackage{geometry}
\usepackage[american]{babel}
\usepackage{amsmath,amsfonts,amsthm,amssymb}
\usepackage[utf8]{inputenc}
\usepackage{csvsimple}
\usepackage{fontenc}
\usepackage{graphicx}
\usepackage{epstopdf}
\usepackage{caption}
\usepackage{exceltex}
\usepackage{ulem}
\usepackage{color}
\usepackage{natbib}
\usepackage{nicefrac}
\usepackage{booktabs}
\usepackage{enumitem}
\usepackage{lscape}
\usepackage{setspace}
\setitemize{leftmargin=*}
\usepackage{hyperref}
\usepackage{eurosym}
\usepackage{multirow}

\date{}
\author{
Malte Jahn\thanks{Department of Mathematics and Statistics, Helmut Schmidt University, 22043 Hamburg, Germany. E-Mail: \href{mailto:jahnma@hsu-hh.de}{\nolinkurl{jahnma@hsu-hh.de}}. ORCID:  \href{https://orcid.org/0000-0002-7165-0428}{0000-0002-7165-0428}.}
}

\title{\textbf{Regressing on distributions: The nonlinear effect of temperature on regional economic growth}}

\begin{document}
\maketitle
\bigskip
\bigskip

\vspace{2cm}
\noindent
\begin{center}
\textbf{Abstract}
\end{center}
A nonlinear regression framework is proposed for time series and panel data for the situation where certain explanatory variables are available at a higher temporal resolution than the dependent variable. The main idea is to use the moments of the empirical distribution of these variables to construct regressors with the correct resolution. As the moments are likely to display nonlinear marginal and interaction effects, an artificial neural network regression function is proposed. The corresponding model operates within the traditional stochastic nonlinear least squares framework. In particular, a numerical Hessian is employed to calculate confidence intervals. The practical usefulness is demonstrated by analyzing the influence of daily temperatures in 260 European NUTS2 regions on the yearly growth of gross value added in these regions in the time period 2000 to 2021. In the particular example, the model allows for an appropriate assessment of regional economic impacts resulting from (future) changes in the regional temperature distribution (mean \emph{and} variance).
\\
\bigskip
\bigskip

\newpage

\section{Introduction}
Identifying the impact of weather and especially, extreme weather, on economic development has been of great interest to economists for many years. In light of current climate change, understanding these impacts becomes even more relevant. 
Previous empirical studies on weather impacts mainly considered linear relationships between weather variables and economic development. The analyses by \citet{Dell12} constitute a standard approach. Their main model (equation 4) is a distributed lag model which implies a linear relationship between a country's growth rate and its current and past temperature levels. Their main finding is that a significant (instantaneous and cumulative) effect of temperature is only present in poor countries, where it is negative.\\
Another standard approach is reflected in the work of \citet{Burke15}. They also consider country-level data to estimate fixed effects models. They include polynomial expressions of temperature to account for nonlinear effects. Their results suggest a global growth-optimal temperature level of 13 degrees Celsius. More recent publications such as \citet{KalkuhlWenz20} also mainly build on the previously discussed econometric approaches. The latter study finds that a temperature increase tends to increase economic growth in cold (world) regions and decrease economic growth in hot (world) regions.\\
In this paper, two main issues regarding the established weather econometric methodologies are addressed. First, the weather variables are parametrized in terms of the moments of their distribution. Corresponding considerations are found in \cite{Hsiang16} but have remained theoretical. Standard approaches \citep{Dell12,Burke15} assume that economic development (in certain time period) depends on the (weighted) average temperature level. However, if we believe that extreme weather situations are particularly relevant for economic growth, the average (i.e.\ the first moment of the distribution) might not adequately describe the relationship. A better representation of the entire temperature distribution in regression models can be achieved by calculating further moments of the distribution, in particular the (central) second moment, the variance. As the influence of the moments is likely nonlinear and might also feature interaction effects, a sensible regression approach needs to allow for such nonlinearities. Here, artificial neural networks (ANN) regression models are suggested which possess the universal approximation property. Regarding the precise estimation strategy, a novel fixed effects ANN model is proposed which is also useful outside the weather econometric context.\\
The second innovation concern the consideration of global and region-specific temperature-growth relationships. Global models obviously assume that the marginal effect of temperature is independent of the location. The term region-specific is often used to describe that the same econometric model is estimated for different types of countries (rich/poor as in \citet{Dell12,Burke15}) or different types of regions (urban/rural as in \citet{HoltermannRische20}) or that the temperature-growth relationship is moderated by a certain variable such as the average income level or temperature level in a region. On the contrary, the ANN approach proposed here allows for a distinct functional relationship between temperature (moments) and economic growth at every location. Precisely, economic growth is expressed as an explicit function of the locations' geographic coordinates.\\
The paper is organized as follows: In the next section, the considered regional weather-economic data set is presented. Then, linear panel models are estimated, implementing the proposed extension regarding the consideration of multiple moments of the temperature distribution as regressors. In section \ref{annsec}, the theoretically more appealing ANN regression model is introduced and applied to the weather-economic data set. The (global) nonlinear marginal effects of temperature on economic growth are derived, including confidence intervals. Section \ref{locann} extends the approach to allow for location-specific marginal effects. The resulting model is illustrated by simulating a scenario where the temperature level increases uniformly by 2 degrees Celsius. The final section concludes and discussed future research directions.

\section{Data set}
\begin{figure}[!h]
\centering
\footnotesize
(a)\hspace{-3ex}\includegraphics[scale=0.52]{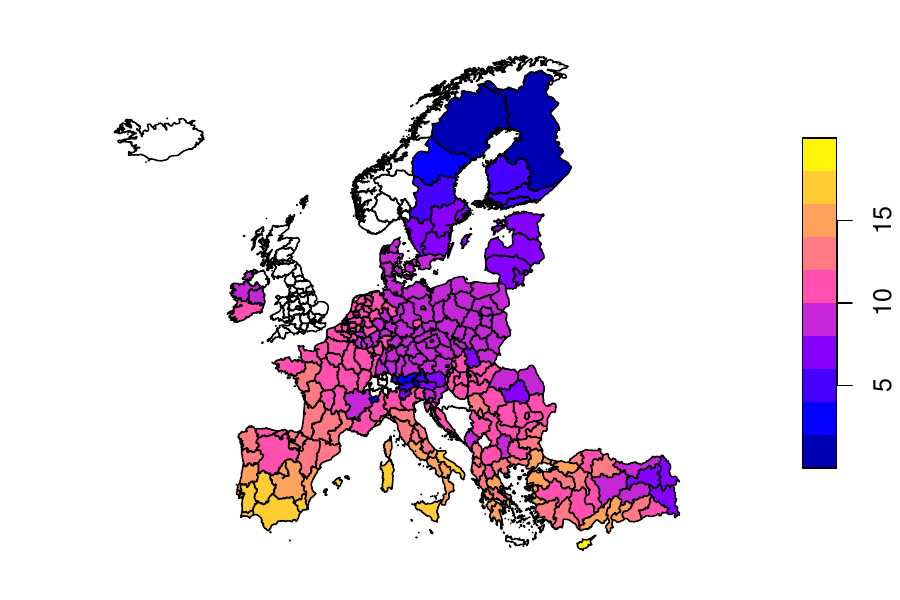}
(b)\hspace{-3ex}\includegraphics[scale=0.52]{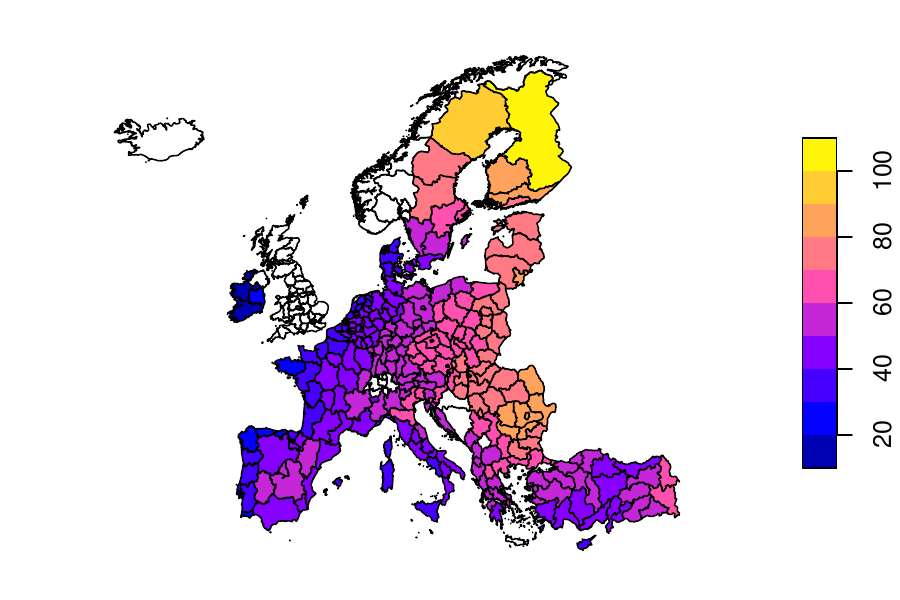}
\caption{\label{tempmap}Mean (a) and variance (b) of the annual distribution of daily mean temperatures}
\end{figure}

The economic data is provided by Eurostat in terms of the real growth rate of the gross value added (GVA) at basic prices under the indicator name `nama\_10r\_2gvagr'. The observations cover the years from 2000 to 2021. The regions correspond to the official European NUTS2 regions (version 2016) and may include countries outside the European Union. Since some regions have missing values for some years and because we are dismissing observations with double digit positive or negative growth rates, the result is an unbalanced panel. Furthermore, a region is required to be observable in at least 5 time periods under the aforementioned restriction. From a total of 260 such regions, we observe 231 on average.\\
The choice of the economic variable is worth a short discussion. Alternative choices may consider the GVA \emph{per capita}, as for example in \citet{Burke15,KalkuhlWenz20,HoltermannRische20}. Both approaches are plausible but the interpretation is slightly different. Using the aggregate GVA, the derived weather impacts may include the relocation of households. This would affect the growth rate of aggregate GVA but not necessarily the growth rate of GVA per capita.\\
The considered weather variable is the daily mean temperature (measured in degrees Celsius) taken from the E-OBS 0.25 gridded dataset (v26.0e). Regarding further information on this meteorologic data, see \citet{Cornes18}. From the gridded (daily) temperature data, regional (daily) averages are calculated first by mapping the grid points from the E-OBS data to the administrative boundaries of the NUTS2 regions. In a second step, the (annual) moments of the distribution of daily mean temperatures are calculated for each region and each year.\\
Figure \ref{tempmap} shows illustrations for the first and second temperature moments with respect to the annual distribution, averaged over all years. While the map for the first moment (temperature mean) is intuitive, the map for the second moment (temperature variance) might require some interpretation. The lowest variances are observed in Ireland which is plausible because of the maritime climate with mild winters and rather cool summers. Some regions in southern Europe such as Sicily display low variances in combination with a high temperature mean, indicating a constantly warm climate throughout the year. On the other extreme, Finnish regions feature high variances due to the large differences between summer and winter temperatures in a generally colder climate.

\section{Linear panel models for weather econometrics}
The assumed setting is that weather and economic variables are observed at multiple locations for multiple time periods (panel data). A common indicator for the economic development in a region is the aggregate income. The real GVA growth rate will be the relevant indicator in this paper but the analysis of weather impacts on other socio-economic variables such as employment, sick leave days, energy consumption etc. would be subject to similar theoretical considerations. The fact that the weather is exogenous (at least in the short/medium run) facilitates the econometric analysis and interpretation. The temporal resolution of the economic data is often much lower (e.g.\ annual) than that of the weather variables (e.g.\ daily). Therefore, the considered weather variables are usually weather indices seeking to describe certain weather features over a given time period in numeric values. It is proposed here that these indices should be the (central) moments of the weather variable distribution in the respective time period. The basic model is\\
\begin{equation}\label{basic}
y_{rt}=\sum_{k=1}^{K}\beta_k m^{(k)}\big(\hat{F}^{temp}_{rt}\big)+ \alpha_{r}+ \mu_{t}+ \epsilon_{rt}.
\end{equation}
The dependent variable $y_{rt}$ is the growth of income (GVA) in region $r$ ($1\leq r \leq R$) in year $t$ ($1 \leq t \leq T$). The expression $m^{(k)}$ ($1\leq k \leq K$) denotes the $k$-th central moment and it is applied to the empirical distribution of (daily) temperatures $\hat{F}^{temp}_{rt}$ in each region and each year. For convenience, the first moment $m^{(1)}$ is defined as the raw moment because the location information of the distribution would otherwise be lost.  In particular, for $K=1$, the model collapses to the usual weather econometric approach \citep{Dell12,Burke15} where (only) the average temperature (first moment) is considered. Furthermore, region-specific effects $\alpha_{r}$ and year-specific effects $\mu_t$ are present. The errors $\epsilon_{rt}$ are assumed to be i.i.d.\ normal with mean zero and standard deviation $\sigma_{\epsilon}$.\\
The inclusion of higher moments allows the model to distinguish between temperature distributions that would be seen as equal if only the first moment was considered. In particular, the impact of a change in the scale or shape of the temperature distribution on economic development can be assessed more accurately. Apart from the novel parameterization of the temperature variables, equation \eqref{basic} constitutes a standard two-way fixed effects model. Let us define the marginal averages $\overline{y}_{r\bullet}=\frac{1}{T}\sum_{t=1}^{T}y_{rt}$, $\overline{y}_{\bullet t}=\frac{1}{R}\sum_{r=1}^{R}y_{rt}$ and $\overline{y}_{\bullet\bullet}=\frac{1}{R\cdot T}\sum_{r=1}^{R}\sum_{t=1}^{T}y_{rt}$, following \citet{Baltagi21}. Using an analogous notation for the regressors $X_{rt}$, three different models can be constructed:

\begin{align}
\label{indfe}y_{rt}-\overline{y}_{r\bullet}&=\left(X_{rt}-\overline{X}_{r\bullet}\right)\beta+ \epsilon_{rt}-\overline{\epsilon}_{r\bullet}\\ 
\label{timefe}y_{rt}-\overline{y}_{\bullet t}&=\left(X_{rt}-\overline{X}_{\bullet t}\right)\beta+ \epsilon_{rt}-\overline{\epsilon}_{\bullet t}\\
\label{twofe}y_{rt}-\overline{y}_{r\bullet}-\overline{y}_{\bullet t}+\overline{y}_{\bullet\bullet}&=\left(X_{rt}-\overline{X}_{r\bullet}-\overline{X}_{\bullet t}+\overline{X}_{\bullet\bullet}\right)\beta+ \epsilon_{rt}-\overline{\epsilon}_{r\bullet}-\overline{\epsilon}_{\bullet t}
\end{align}

Model \eqref{indfe} is the usual one-way fixed effects model and it is appropriate when there are no time-specific effects ($\mu_{t}=0$ for all $t$). On the other hand, model \eqref{timefe} is appropriate when there are no region-specific effects ($\alpha_{r}=0$ for all $r$). Finally, model \eqref{twofe} is the two-way fixed effects model including both effects. Note that the overall variable means $\overline{y}_{\bullet\bullet}$ and $\overline{X}_{\bullet\bullet}$ need to be part of the transformation for the two-way model \citep[][eq. 3.5]{Baltagi21}. The benefit of the three within-transformed model formulations is that they can be estimated by ordinary least squares (OLS) without the need to consider $\alpha_{r}$ and/or $\mu_t$ as explicit model parameters. This is known as the equivalence of the fixed effects and the least squares dummy variable (LSDV) approach. The corresponding reduction in the computational effort will be especially relevant for the ANN regression model discussed later. The transformed error terms imply that the sum of sample errors in the within-dimension(s) must be zero. These can be understood as constraints on the errors with a reduction in the degrees of freedom equivalent to the number of parameters.\\
The results of the estimation of the linear panel models for the weather-economic data set described in the previous section are shown in Table \ref{linear}. In addition to the three true panel models, the pooled model is included which assumes that there are neither region-specific nor time-specific effects. Three performance measures are used to compare the models, the Akaike information criterion (AIC), the Bayesian information criterion (BIC) and the (unbiased) estimate of the error standard deviation. All of these measures include the respective degrees of freedom which can be relatively large, $R+2$ for model \eqref{indfe}, $T+2$ for \eqref{timefe} and $R+T+1$ for \eqref{twofe}. Correspondingly, the BIC prefers the smaller \eqref{timefe} over the larger and AIC-optimal \eqref{twofe}. The unbiased estimate of $\sigma_{\epsilon}$ is used to construct the standard errors and yields a directly interpretable measure of the performance.\\
Note that in the two-way FE model, both coefficients are insignificant, indicating the absence of a \textit{linear} relationship between the temperature moments and economic growth. As pointed out in the introduction, a nonlinear relationship is, in fact, quite plausible. The usual approach to deal with this inside the linear panel framework is to include higher-order polynomials of the regressors. This is done, e.g. in \citet{Burke15,KalkuhlWenz20}, where second-order polynomials of temperature and precipitation averages (first moments) are included.\\
When several moments of a weather variable are already present as proposed in \eqref{basic}, including polynomials for these moments increases the model size significantly. To capture possible interaction effects, even more regressors would have to be constructed. At this point, interpretation and model selection can already become a challenge because the number of (possible) interaction terms grows quickly with the order of the involved moments and polynomials. Furthermore, other types of interaction effects such as region-specific temperature impacts would make the model even larger and more difficult to work with. Therefore, an alternative ANN regression approach is presented in the following section.\\

\begin{table}[!h]
  \centering
  \begin{tiny}
    \begin{tabular}{ccccc}
    \toprule
      coefficient&pooled& region FE & time FE & two-way FE\\
      \midrule
	$m^{(1)}(\hat{F}^{temp})$&-0.0407&-0.4196 &-0.0223 &-0.0362 \\
									&(0.0170)&(0.0730)&(0.0134)&(0.0668)\\
	$m^{(2)}(\hat{F}^{temp})$&-0.0221 &0.0093 &0.0227 &0.0046  \\
									&(0.0028)&(0.0049)&(0.0024)&(0.0047)\\
	\midrule
	model $df$&3&262&24&283\\
	AIC&26379&26135&23857&22934\\
   	BIC&26399&27846&24013&24783\\
	$\sigma_{\epsilon}$&3.247&3.092&2.528&2.252\\
	\bottomrule
    \end{tabular}
    \end{tiny}
     \caption{Linear panel models for the effect of temperature on economic growth}
    \label{linear}
\end{table}

\section{Fixed effects artificial neural network regression}\label{annsec}
\begin{figure}[!h]
\centering
\footnotesize
\includegraphics[scale=0.4]{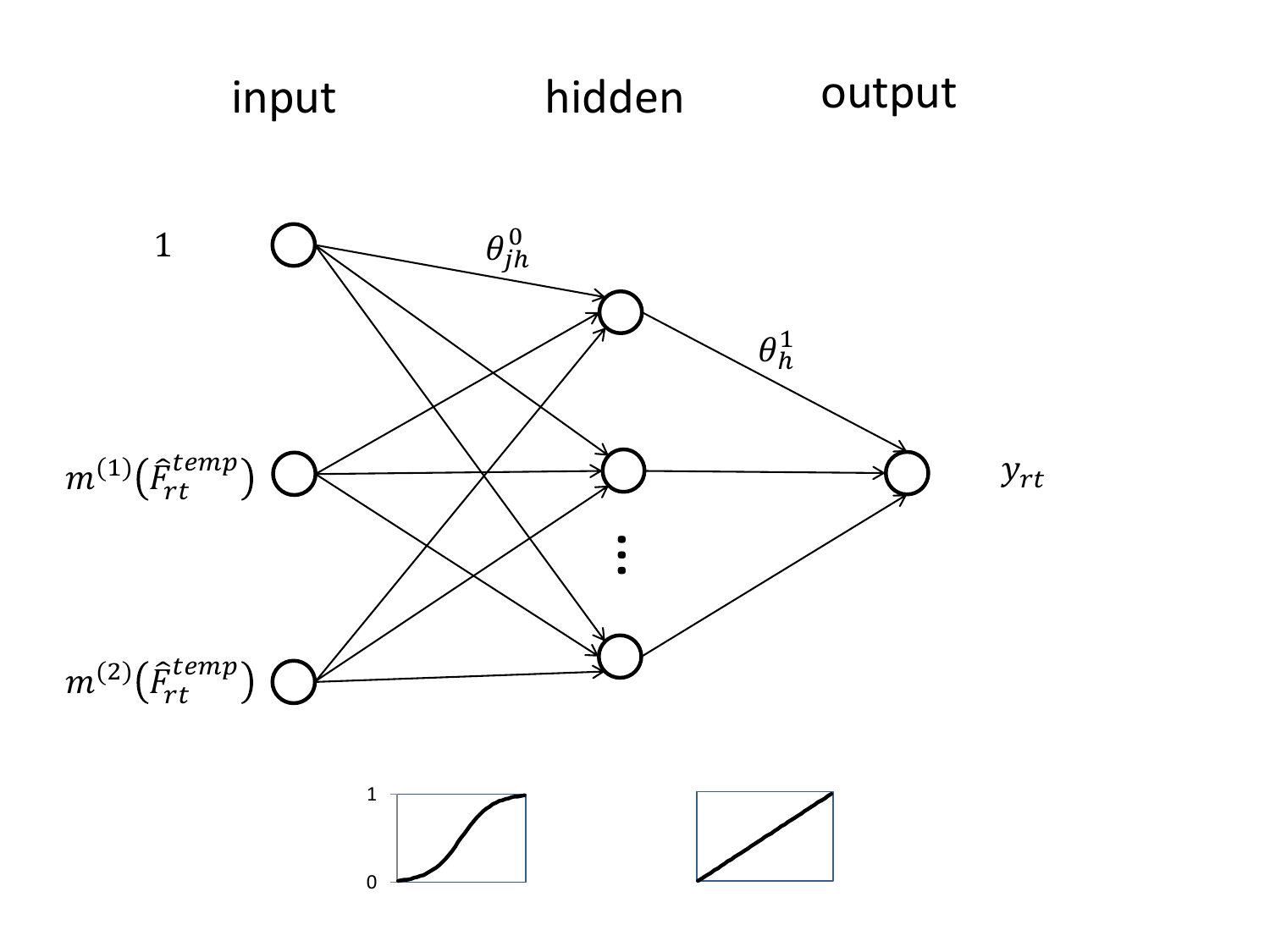}
\caption{\label{netvisual}Network graph for the pooled ANN regression model}
\end{figure}

Artificial neural networks are powerful tools for nonlinear regression and have been applied to various problems, including economic problems such as the analysis of economic growth \citep[e.g.][]{FengZhang14, Jahn20}. The ANN regression approach allows for an (almost) arbitrary influence of the regressors on the dependent variable. This flexibility of the ANN regression approach follows from the universal approximation theorem which states that any continuous regression function can be approximated arbitrarily close on a compact set by a network of sufficient complexity \citep[cf.][p. 616]{KockTeraesvirta14}.\\
The ANN regression function employed here is precisely a single hidden layer feedforward network  (SLFN) with sigmoid activation of the hidden layer (logistic function $\phi$) and linear activation of the output layer. The restriction to one hidden layer is without loss of generality since one hidden layer is enough to guarantee the universal approximation property. For vector-valued argument $x$ and parameter vector $\theta=(\theta^0,\theta^1)$, the scalar output is defined as
\begin{equation}\label{annreg}
f^{ANN}(\theta^0,\theta^1,x)=\sum_{h=1}^{H}\theta^1_{h}\cdot \phi\! \left(\sum_{j=1}^{J}\theta^0_{jh} x_j\right).
\end{equation}
To ensure identifiability, it is assumed that the network is fully-connected in the sense that all $\theta^1_{h}$ are different from zero. The index $h$ ($1 \leq h \leq H$) refers to the neurons in the hidden layer and $j$ ($1 \leq j \leq J$) refers to the regressor variables which are called inputs in the context of ANN.  The number of hidden neurons $H$ governs the complexity and is chosen by model selection procedures. In practice, the goal is to find a compromise between the ability to describe the assumed arbitrary nonlinear relationship (which increases with $H$) and the number of parameters which is given by $(J+1)H$.\\
Turning to the econometric problem considered in this paper, Figure \ref{netvisual} shows the network graph for the ANN model without any region- or time-specific effects, i.e., the pooled version of \eqref{basic}. Explicit dummy variables for region- or time-specific effects would appear as additional neurons in the input layer. This, in turn, would drive up the number of parameters even more than in the linear panel models. Therefore, it makes even more sense to estimate the ANN model on the within-transformed data to get rid of the dummies for the numerical estimation. However, it also means that the sample errors do not exactly sum to zero in the within-dimension(s) as we transition to ANN. In the present application, the deviations from zero are negligibly small but this not guaranteed. For each of the four models considered in Table \ref{linear}, an ANN analogue is constructed by replacing the linear expression $x\beta$ by the nonlinear function $f^{ANN}(\theta,x)$. This yields a nonlinear least squares problem and appropriate algorithms can be used to minimize the sum of squared errors numerically on the whole sample, similarly to \cite{Jahn20}. The goal is to find the global minimum with respect to the parameters as in traditional nonlinear least squares problems. The benefit of employing ANN in the traditional stochastic framework is that we can use the numerical Hessian to approximate the covariance matrix. This, in turn, allows for the calculation of confidence intervals for predictions which greatly improves the interpretability of the model. Another consequence of the stochastic framework is that model selection can be done on the basis of information criteria without the need for an explicit validation set. 

\begin{table}[!h]
  \centering
  \begin{tiny}
       \begin{tabular}{cccccc}
    \toprule
      model&$H$&model $df$& AIC & BIC & $\sigma_{\epsilon}$\\
      \midrule
	pooled&3&12&26299&26377&3.218\\
	pooled&8&32&26242&26451&3.194\\
	\midrule
	ind. FE&4&272&25981&27758&3.043\\
	ind. FE&9&287&25956&27831&3.031\\
	\midrule
	time FE&3&31&23757&\textbf{23959}&2.501\\
	time FE&6&40&23724&23985&2.491\\
	\midrule
	two-way FE&2&287&22899&24794&2.243\\
	two-way FE&10&311&\textbf{22840}&24871&2.225\\
	\bottomrule
    \end{tabular}
    \end{tiny}
     \caption{ANN panel models for the effect of temperature on economic growth}
    \label{nonlinear}
\end{table}

\begin{figure}[!h]
\centering
\footnotesize
(a)\hspace{-3ex}\includegraphics[scale=0.38]{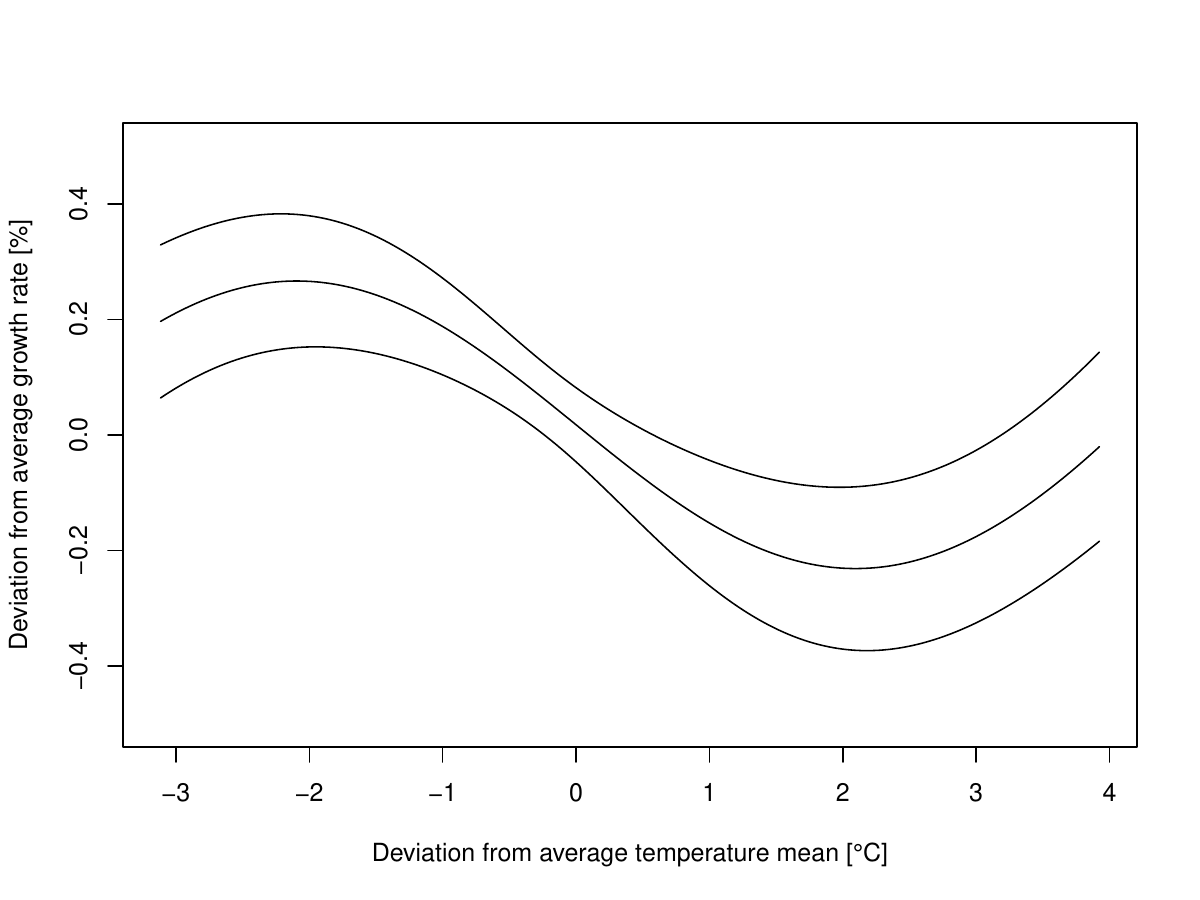}
(b)\hspace{-3ex}\includegraphics[scale=0.38]{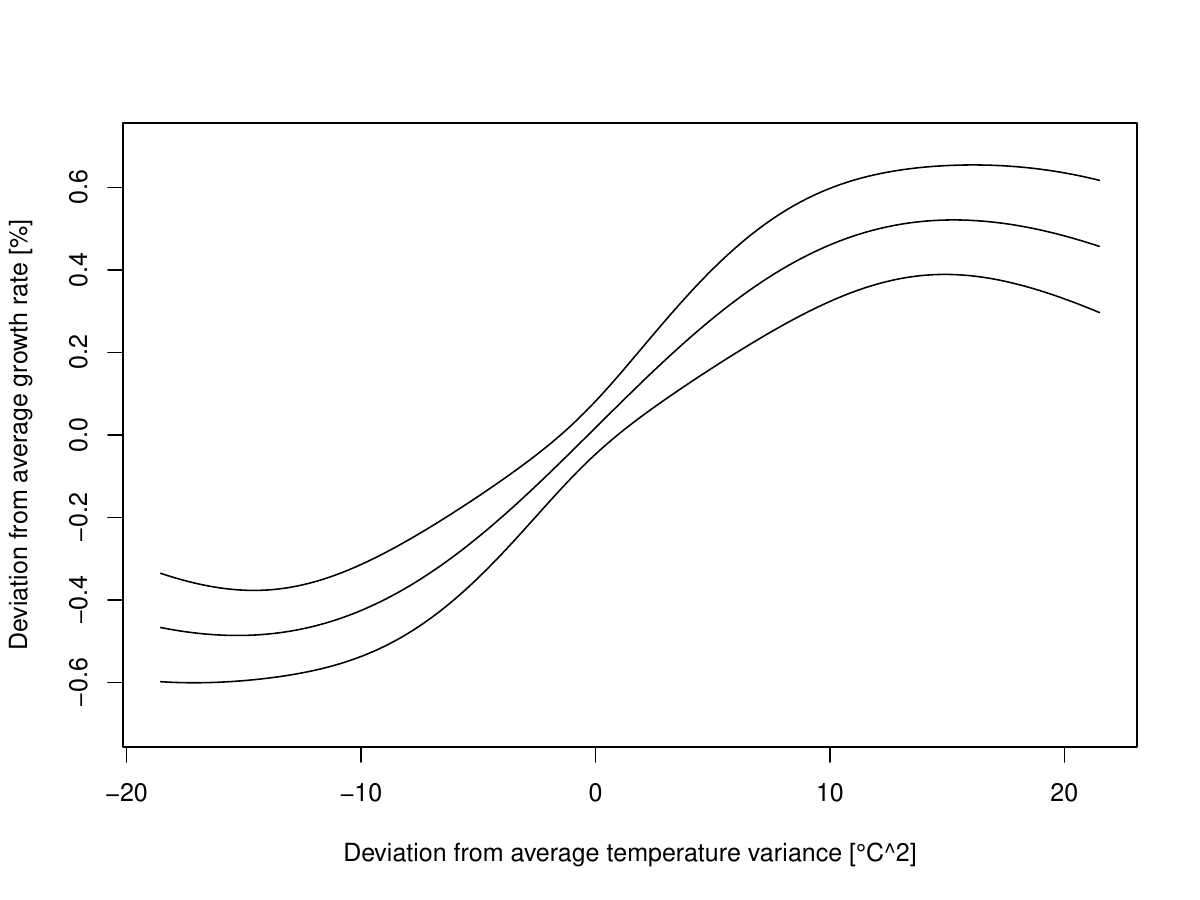}
(c)\hspace{-3ex}\includegraphics[scale=0.38]{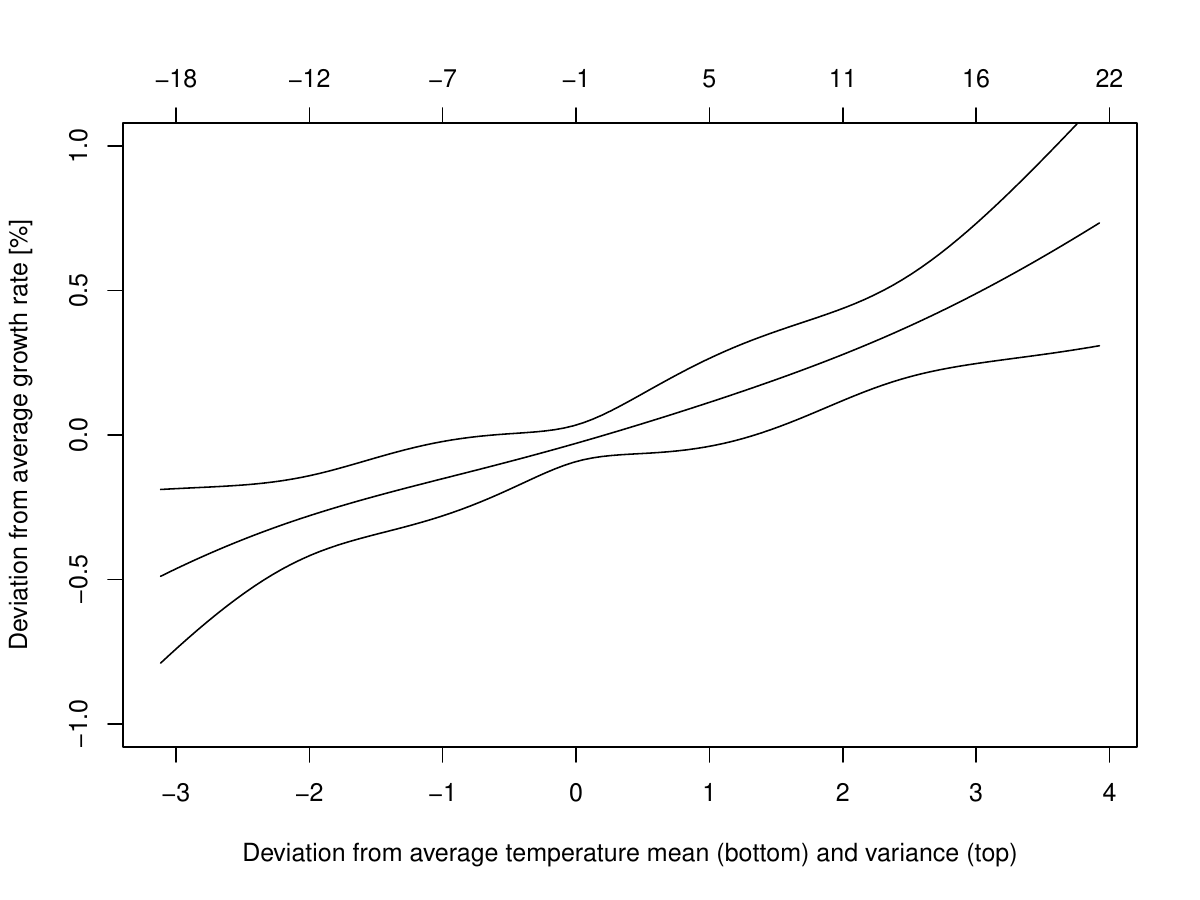}
(d)\hspace{-3ex}\includegraphics[scale=0.38]{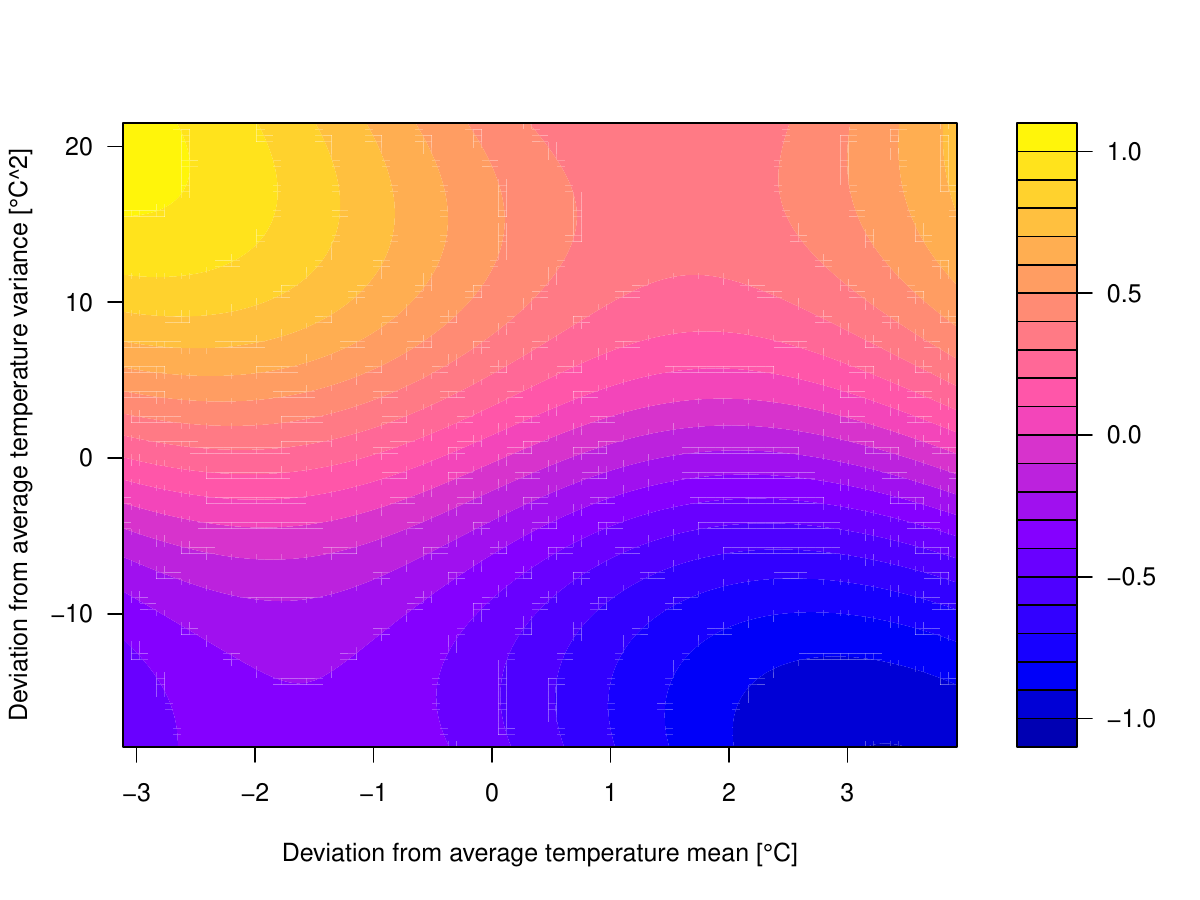}
\caption{\label{margplot}Illustration of the marginal effect at the mean for the first (a) and second (b) temperature moment,  interaction effect (c) and contour plot (d). Pointwise 95\% confidence intervals added to (a)-(c).}
\end{figure}

The estimation results for the weather-economic data for different $H$ are shown in Table \ref{nonlinear}. The displayed choices for $H$ correspond to the AIC- and BIC-optimal specifications.  First note that both the AIC-optimal model (two-way FE, $H=10$) and the BIC-optimal model (time FE, $H=3$) are superior to the corresponding optimal linear models from Table \ref{linear}. Since the ANN models have a lot of parameters, the two information criteria yield very different optimal values for $H$ for the same FE specification. Here, the goal is to extract the general functional relationship between the regressors and the dependent variable (and not to predict or forecast). Therefore, simpler models are especially relevant and the BIC-optimal model is chosen for further analysis.\\
Regarding the discussion of the results, the parameter vector $\theta$ itself does not have a direct interpretation. To visualize the marginal effect of the temperature variables, $f^{ANN}(\hat{\theta},x)$ is computed for different $x$ keeping all but one variable constant at their mean values. This is known as marginal effects at the mean and is a standard procedure for nonlinear regression models. Confidence intervals are calculated from the formula for the conditional variance  $Var(f(x,\hat{\theta}))=\sigma_{\epsilon}^2 \nabla f(x,\hat{\theta})^{T}\mathcal{H}^{-1}\nabla f(x,\hat{\theta})$, see \citet{RivalsPersonnaz00} which holds for general (nonlinear) regression functions $f$. It is a generalization of the well-known formula for the OLS model where the gradient is simply given by $\nabla x\beta=x$ and the Hessian is $\mathcal{H}=X^{T}X$. For neural network regression functions, the gradient $\nabla f^{ANN}(\hat{\theta},x)$ can be calculated layer-wise which is known as backpropagation and the Hessian $\mathcal{H}$ is approximated numerically during or after the optimization. Figure \ref{margplot} displays illustrations for the marginal effects of the first and second temperature moment as suggested by the BIC-optimal model (time FE, $H=3$), including (approximate) 95\% confidence intervals. First note the axes of the plot. Due to the within-transformation, the involved variables have to be interpreted as deviations from their mean value (in the within-dimension(s)). Subfigure (a) implies that, holding the temperature variance constant, an increase in the temperature level (positive deviation) would result in lower economic growth rates. This is in line with the generally negative coefficients for the temperature mean in the linear panel models (Table \ref{linear}). However, the ANN model displays decreasing marginal effects for more extreme deviations. It can be concluded that the regional economies are more robust to extreme temperature level deviations than a linear relationship would suggest. The new element in this paper is to consider the temperature variance and subfigure (b) indicates that higher variances are significantly better for economic growth (holding the temperature level constant). This is a novel empirical result, so further research is needed to understand why that is the case. A possible interpretation could be made on the basis that an unusually high variance (in combination with a `usual' mean, as the latter is kept constant) describes the consequence of a very cold winter in combination with a very hot summer. This would most certainly affect the energy sector through increased demand for heating/cooling. Although negative effects on other sectors might be working in an opposite direction, increased activity of the energy sector could give a theoretical explanation for the observed relationship. Subfigure (c) considers the simultaneous change in mean and variance and it can be concluded that the positive variance effect dominates. The most comprehensive illustration is the contour plot from Figure \ref{margplot}~(d). The previous subfigures can be interpreted as slices through this plot (horizontally at zero for (a), vertically at zero for (b) and diagonally for (c)). An important conclusion is that the economic effect of warming (increase in the temperature level) is largely moderated by the temperature variance. This justifies the proposed method of including higher moments of the temperature distribution into the analysis. Regarding the practical implications, the novel approach allows for a more accurate assessment of the economic effects of climate change, assuming that regional information on the change in temperature variability are available.

\section{Location-specific growth impacts of temperature} \label{locann}
The previously considered linear and ANN panel models are global models in the sense that the functional relationship between the temperature moments and the economic growth rate is assumed to be the same at every location. As mentioned in the introduction, the term region-specific is often used in the literature to describe that the same econometric model is estimated for different types of countries (rich/poor) as in \citet{Dell12,Burke15}) or different types of regions (urban/rural) as in \citet{HoltermannRische20}). The ANN framework allows us to construct models where the functional relationship is different at each geographic location. The idea is to include the geographic coordinates ($lat$, $lon$) of the centroids of the regions as additional regressor variables (input neurons). The ANN model not only allows for a nonlinear effect of these variables (to filter out location-specific growth levels) but, in particular, for arbitrary interaction effects with the temperature variables. In other words, the marginal effects of the temperature moments depend on the geographic coordinates.\\
As they are constant for each region, the coordinates would be eliminated by any within-transformation which assumes region-specific levels $\alpha_r$. Therefore, if the goal is to estimate region-specific impact functions, the basis should be the pooled model or the time FE model. For the present data, the time FE model was preferred among the global models, so it is natural to consider this model for the inclusion of location-specific effects.  Table \ref{location} reveals that location-specific effects are significant in the sense that a large drop in AIC and BIC is obtained. The BIC-optimal network complexity increases from $H=3$ for the global model to $H=6$ for the location-specific model. This is plausible because economic growth is expected to be a rather complex function of the geographic coordinates.

\begin{table}[!h]
  \centering
  \begin{footnotesize}
    \begin{tabular}{cccccc}
    \toprule
      model&$H$&$df$& AIC & BIC & $\sigma_{\epsilon}$\\
      \midrule
	time FE&3&31&23757&23959&2.501\\
	local time FE&6&52&23112&23452&2.343\\
	\bottomrule
    \end{tabular}
    \end{footnotesize}
         \caption{Location-specific ANN panel model with time FE}
    \label{location}
\end{table}

\begin{figure}[!h]
\centering
\footnotesize
\includegraphics[scale=0.5]{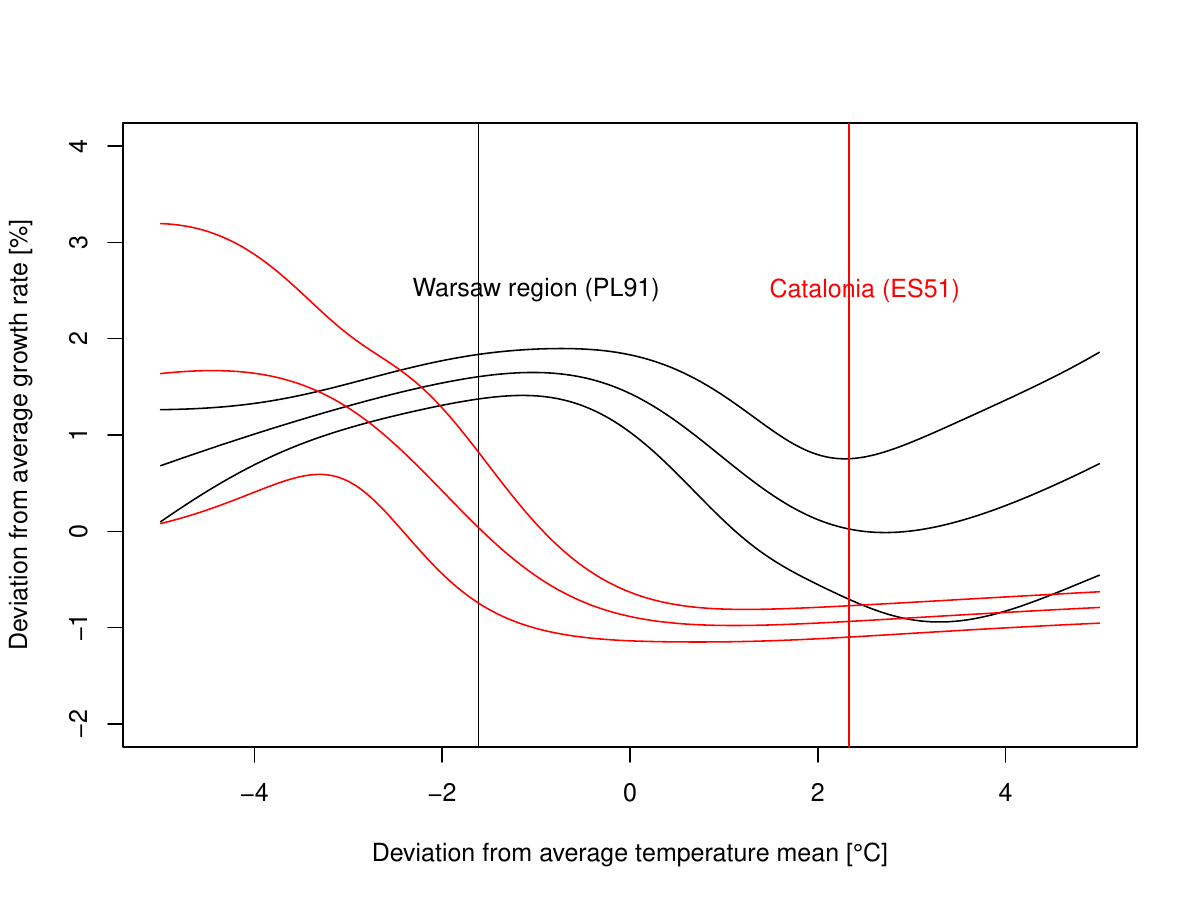}
         \caption{Marginal effect at the mean for the first temperature moment at two locations including 95\% confidence intervals}
    \label{locmarg}
\end{figure}

In order to illustrate the results, marginal effects of the two temperature moments (and associated confidence intervals) can be illustrated the same way as before. Obviously, the corresponding curves will now look different at each location. Figure \ref{locmarg} displays the marginal effect of the first temperature moment (the mean) according to the location-specific time FE model at Warsaw (code PL91, black) which is located in central Poland and Catalonia (Code ES51, red) which is located at the Spanish Mediterranean coast. The vertical lines indicate the temperature means which are usually observed in these regions (cf. Figure \ref{tempmap}(a)). The model suggests that the temperature level is almost optimal for economic growth in Warsaw. Only an extreme increase in the temperature mean of several degrees Celsius would significantly affect economic growth. For Catalonia, it is first observed that the curve is generally below that of Warsaw. This shows that the geographic coordinate variables also control for level effects as one would assume in a linear model. However, the main interest lies in the shape of the curve. The curve for Catalonia is also nearly constant around the position of the vertical line, indicating a relative robust economy with respect to the temperature mean. Catalonia would theoretically benefit from a temperature level as low as that of Warsaw, which is obviously unrealistic. This is also reflected in the widening of the confidence interval in that region.

\begin{figure}[!h]
\centering
\footnotesize
\includegraphics[scale=0.75]{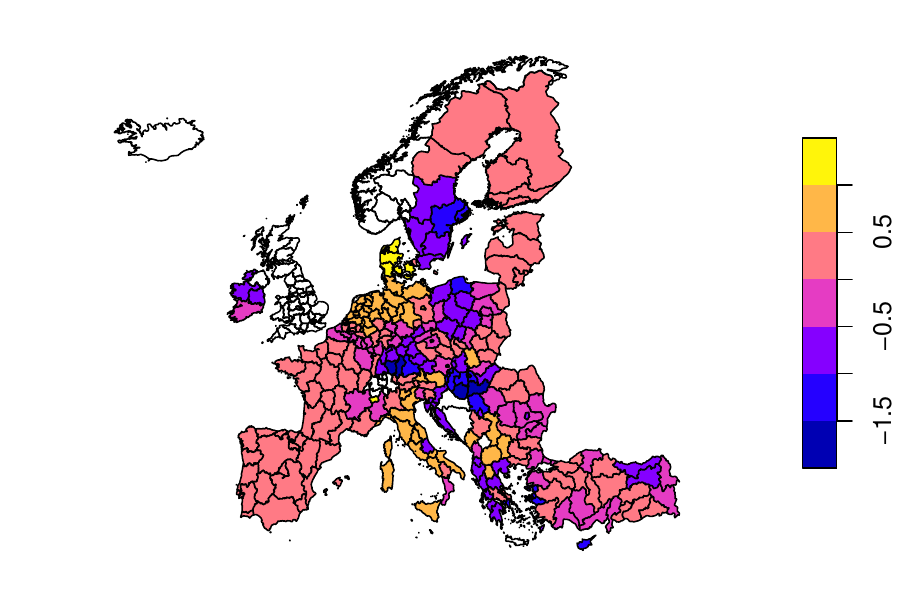}
\caption{Scenario: 2 degree Celsius uniform temperature mean increase}
\label{scenmap}
\end{figure}

In order to provide a more comprehensive understanding of the location-specific model (without showing 260 marginal effect curves), a temperature scenario is considered for all regions. It is assumed that the temperature mean increases uniformly by 2 degrees Celsius and the variance remains constant. Figure \ref{scenmap} shows the effect on the economic growth rate resulting from the warming scenario according the location-specific ANN model. Most regions would experience a moderate change in the growth rate by +/-0.5 percentage points, for example in France, Spain or Portugal. This implies that the regional economies there are relatively robust against a warming in the form of a shift in the mean of the temperature distribution. This is line with Figure \ref{margplot}(a) from the global model and with Figure \ref{locmarg} from the location-specific model. More significant losses can be observed, e.g., in southern Germany and Hungary. On the other hand, there are also regions which would experience positive economic effects, the largest are found in Denmark. In general, positive economic effects from an increase in the temperature level have also been reported in other studies \citep[][for very cold world regions]{KalkuhlWenz20}. However, further research is needed to understand the underlying impact channels in the present case. For example, one could speculate that the summer tourism and the agricultural output in Danish regions would benefit from a higher temperature level.

\section{Conclusion and outlook}
The general contribution of the paper is to provide a framework for incorporating the distribution information of explanatory variables in a time series or panel setting where the explanatory are available at a higher temporal resolution than the dependent variable by including (central) moments of the explanatory variables as regressors. An important area of application are climate econometric problems where the goal is to estimate a relation between (daily) weather variables and (annual) economic variables. The analysis of the economic growth rates in European regions revealed that corresponding nonlinear marginal and interaction effects for the temperature moments can be addressed by an artificial neural network regression function. The ANN regression function is considered as part of the traditional nonlinear least squares framework which facilitates the interpretation by allowing, i.a., for the calculation of confidence intervals. The potential of the ANN regression function can be exploited even further to model location-specific marginal effects. Including the geographic coordinates as regressors not only allows  `controlling' for location-specific level effects but, in particular, for describing marginal effects of the other regressors (here: the temperature moments) as explicit functions of the geographic coordinates.\\
The directions for future research are manifold. For example, it might be informative to consider moments calculated from daily maximum or minimum temperatures instead of or in addition to the daily mean temperature.\\
Furthermore, it seems obvious to include additional weather variables into the analysis, most importantly precipitation. This is done, e.g., in \citet{Burke15,KalkuhlWenz20}. Again, the expected (nonlinear) interaction effects between temperature and precipitation moments could be addressed convincingly by the ANN regression model. The main panel model in \citet{KalkuhlWenz20} also provide an alternative regarding the choice of regressors. The inclusion of first differences of temperature (and precipitation) in addition to the levels allows for a distinction between short- and long-run growth effects.\\
An aspect from the analysis of \citet{HoltermannRische20} is that they account for explicit spatial correlation structures in the dependent variable. This could also be incorporated into the ANN regression framework.\\
Finally, it seems worth investigating whether actual climate projections from regional climate models can be aligned with the proposed econometric model. In the best case, it would be possible to provide improved predictions for regional economic impacts of climate change.\\

\textbf{Acknowledgment}\\
I acknowledge the E-OBS dataset from the EU-FP6 project UERRA (https://www.uerra.eu) and the Copernicus Climate Change Service, and the data providers in the ECA\&D project (https://www.ecad.eu)\\


\begin{thebibliography}{}
 \bibitem[Baltagi(2021)]{Baltagi21}Baltagi B.H. (2021): Econometric Analysis of Panel Data. Sixth Edition. Springer Texts in Business and Economics. Springer, Cham. 
 \bibitem[Burke et al.(2015)]{Burke15}Burke M., Hsiang S.M., Miguel E. (2015): Global nonlinear effect of temperature on economic production, \emph{Nature} 527.
\bibitem[Cornes et al.(2018)]{Cornes18}Cornes R., van der Schrier G., van den Besselaar E.J.M., Jones P.D. (2018): An Ensemble Version of the E-OBS Temperature and Precipitation Datasets, \emph{J. Geophys. Res. Atmos.} 123.
\bibitem[Dell et al.(2012)]{Dell12}Dell M., Jones B.F., Olken B.A. (2012): Temperature Shocks and Economic Growth: Evidence from the Last Half Century, \emph{American Economic Journal: Macroeconomics} 4(3), 66-95.
\bibitem[Feng and Zhang(2014)]{FengZhang14} Feng L., Zhang J. (2014): Application of artificial neural networks in tendency
forecasting of economic growth, \emph{Economic Modelling} 40, 76-80.
\bibitem[Holtermann and Rische(2020)]{HoltermannRische20} Holtermann L., Rische M.C. (2020): The Subnational Effects of Temperature on Economic Production: A Disaggregated Analysis in European Regions, \emph{Working paper}, \mbox{https://www.researchgate.net/publication/347959854}.
\bibitem[Hsiang(2016)]{Hsiang16}Hsiang, S. (2016): Climate econometrics. \emph{Annual Review of Resource Economics} 8, 43-75.
\bibitem[Jahn(2020)]{Jahn20}Jahn M. (2020): Artificial neural network regression models: Predicting economic growth, \emph{Economic Modelling} 91, 148-154.
\bibitem[Kalkuhl and Wenz(2020)]{KalkuhlWenz20}Kalkuhl, M., Wenz, L. (2020): The impact of climate conditions on economic production. Evidence from a global panel of regions. \emph{Journal of Environmental Economics and Management} 103, 102360.
\bibitem[Kock and Teräsvirta(2014)]{KockTeraesvirta14}Kock, A.B., Teräsvirta, T. (2014): Forecasting performances of three automated modelling techniques during the economic crisis 2007-2009. \emph{International Journal of Forecasting} 30, 616-631.
\bibitem[Rivals and Personnaz(2000)]{RivalsPersonnaz00} Rivals, I., Personnaz, L. (2000): Construction of confidence intervals for neural networks based on least squares estimation. \emph{Neural Networks} 13, 463-484.
\end{thebibliography}
\end{document}